\documentclass{iau}
\usepackage{graphicx} 

\title[IAUS291.~~Proper Motions of Magnetars] 
{Near IR Astrometry of Magnetars} 

\author[S. P. Tendulkar]  
{Shriharsh P. Tendulkar}

\affiliation{Department of Astronomy and Astrophysics, California Institute of Technology \\ email: {\tt spt@astro.caltech.edu}}

\pubyear{2012}
\volume{291}  
\jname{\mbox{Neutron Stars and Pulsars: Challenges and Opportunities after 80 years}}
\editors{J. van Leeuwen, ed.} 
\begin{document}

\maketitle

\begin{abstract}

We report on the progress of our five-year program for astrometric
monitoring of magnetars using high-resolution NIR observations using
the laser guide star adaptive optics (LGS-AO) supported NIRC2 camera
on the 10-meter Keck telescope. We have measured the proper motion of
two of the youngest magnetars, SGR\,1806$-$20 and SGR\,1900$+$14,
which have counterparts with K $\sim$21\,mag, and have placed a preliminary upper limit on the motion of the young AXP\,1E\,1841$-$045. The precision of the proper motion measurement is at the milliarcsecond per year level. Our proper motion measurements now provide evidence to link SGR\,1806$-$20 and SGR\,1900$+$14 with neighboring young star clusters.  At the distances of these magnetars, their proper motion corresponds to transverse space velocities of $350\pm100\,\mathrm{km\,s^{-1}}$  and $130\pm30\,\mathrm{km\,s^{-1}}$ respectively. The upper limit on the proper motion of AXP\,1E\,1841$-$045 is $160\,\mathrm{km\,s^{-1}}$. With the sample of proper motions available, we conclude that the kinematics of the magnetar family are not distinct from that of pulsars.
 
\end{abstract}

 \setlength\fboxsep{0pt} 
\setlength\fboxrule{0.5pt}

\firstsection 
\section{Introduction}
Magnetars or highly magnetized neutron stars were proposed by Thompson \& Duncan (1996) as a unified model to explain the phenomena of soft gamma repeaters (SGRs) and anomalous X-ray pulsars (AXPs). The short and intense $\gamma$-ray flares were attributed to violent reconnections of a twisted magnetic field and the anomalously high quiescent X-ray emission from AXPs was ascribed to the decay of the same strong magnetic field. Indeed, the detection of a strong SGR-like flare and rotation glitch in the AXP 1E\,2259$+$586 by Kaspi et al. (2003) was a strong validation of the unified nature of SGRs and AXPs. 

Magnetars are currently a small family with only 9 SGRs and 12 AXPs identified from X-ray data and $\gamma$-ray bursts. Of these only 2 SGRs and 4 AXPs have well identified near infrared (NIR) counterparts. Identifying NIR counterparts for magnetars and comparing their NIR emission to X-ray emission allows us to constrain their emission mechanisms. The precise localization in NIR bands also offers an excellent avenue for the measurement of proper motions of magnetars thus allowing us to identify their birth sites and estimate kinematic ages. Here we report on the progress of our campaign to for astrometric and photometric measurements of magnetars using high-resolution NIR imaging.

\section{Observations \& Results}
The NIRC2 camera at the 10-meter Keck II telescope is designed to utilize the high resolution ($\sim 6$\,milli-arcseconds) achieved by the Laser Guide Star Adaptive Optics (LGS-AO) system.
 We used the $10 \times 10^{\prime\prime}$  ``narrow'' mode of the NIRC2 camera to observe the targets over multiple epochs from 2005 till 2010. After flat-fielding and dark subtracting the images, we corrected for the instrumental distortion of NIRC2 by a polynomial transformation. In order to reduce systematic errors caused due to residual distortion, we registered each target field at the same position on the detector in each epoch. We calculated the proper motion of each star with respect to a grid of neighboring stars using the optimal weighting scheme developed in Cameron et al.  (2009). In this scheme, we chose optimal weights of each star-target vector by accounting for position jitter correlations (tip-tilt anisoplanatism). The relative astrometry was corrected for the bulk motion of the field by modeling the Galactic rotation curve.

\begin{figure}[tb]
  \begin{center}%
    \fbox{\includegraphics[width=0.42\textwidth]{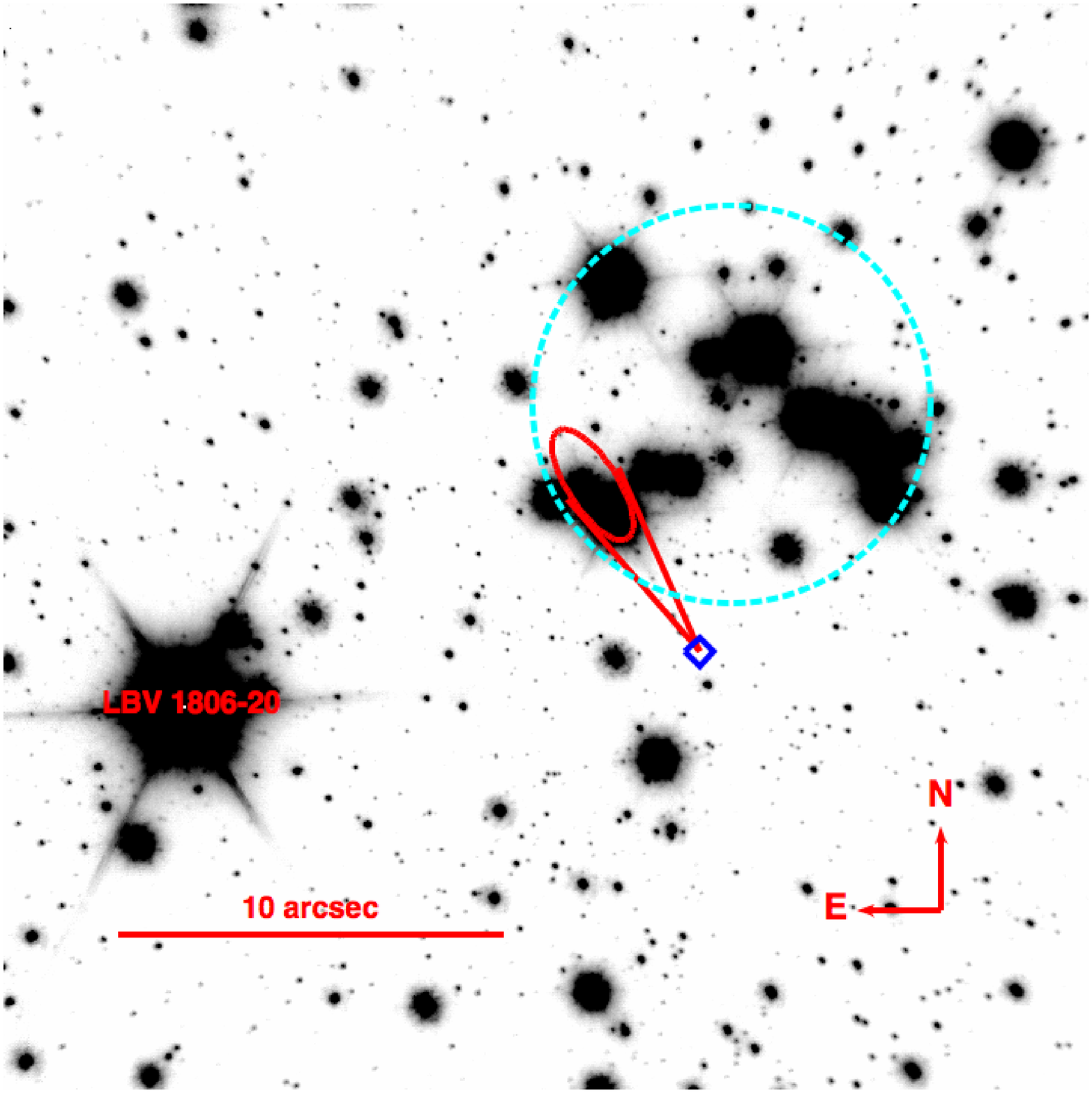}}
    \fbox{\includegraphics[width=0.42\textwidth]{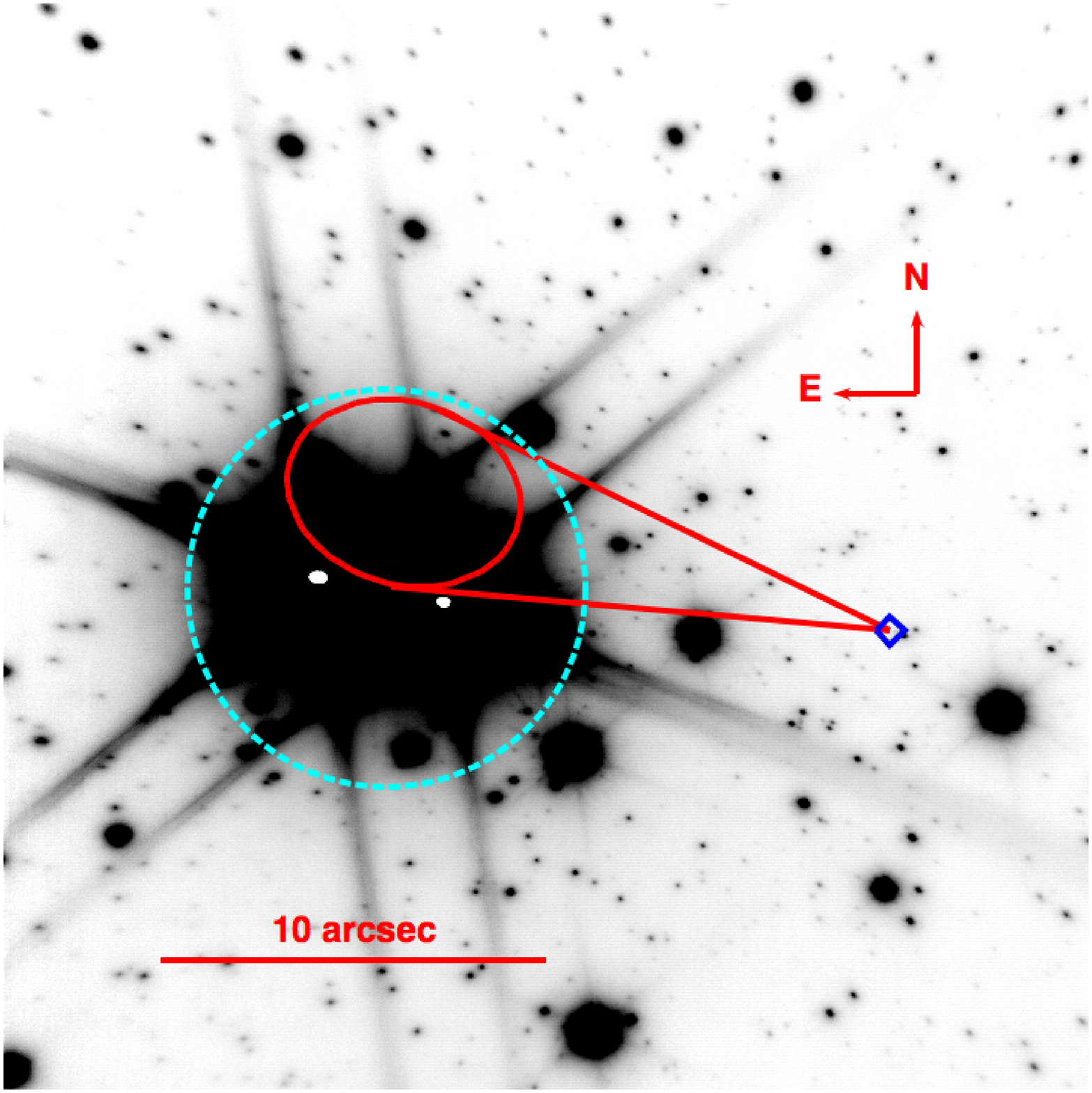}}
    \caption{\textit{Left Panel:} The motion of SGR\,1806$-$20 (blue diamond) traced backwards in time by 650\,yrs is denoted by the red ellipse, coincident with the cluster of massive stars (dashed cyan circle) identified by Fuchs et al. (1999). \textit{Right Panel:} The motion of SGR\,1900$+$14 (blue diamond) traced backwards in time by 6\,kyr is marked by the red ellipse, also coincident with the cluster of massive stars (dashed cyan circle) identified by Vrba et al. (2000). Figures adapted from Tendulkar et al. (2012).} 
    \label{fig:sgr_proper_motions}
  \end{center}
\end{figure}

From this astrometric program, we reported the proper motions of SGR\,1806$-$20 and SGR\,1900$+$14 to be $350\pm100\,\mathrm{km\,s^{-1}}$  and $130\pm30\,\mathrm{km\,s^{-1}}$~(Tendulkar et al. 2012). Figure~\ref{fig:sgr_proper_motions} shows the direction of motion of SGR\,1806$-$20 and SGR\,1900$+$14 with respect to their neighboring objects. The motion of both the magnetars can be traced backwards to the clusters of massive stars with which the magnetars were long associated (Fuchs et al. 1999, Vrba et al. 2000), providing a model-independent kinematic age of $\sim650\,$yr and 6\,kyr respectively.

\subsection{AXP\,1E\,1841$-$045}
Testa et al. (2008) proposed a NIR counterpart for AXP\,1E\,1841$-$045 but the identification was not conclusive. We observed AXP\,1841$-$045 using the LGS-AO and the NIRC2 camera on 9 epochs between 2005 and 2009. The coadded exposure time between 15 to 45 minutes at each epoch depending on the observing circumstances. The limiting magnitude for each coadded observation was $K_s \approx 21\,$mag. 
 
\begin{figure}[tb]
  \begin{center}
    \fbox{\includegraphics[width=0.42\textwidth]{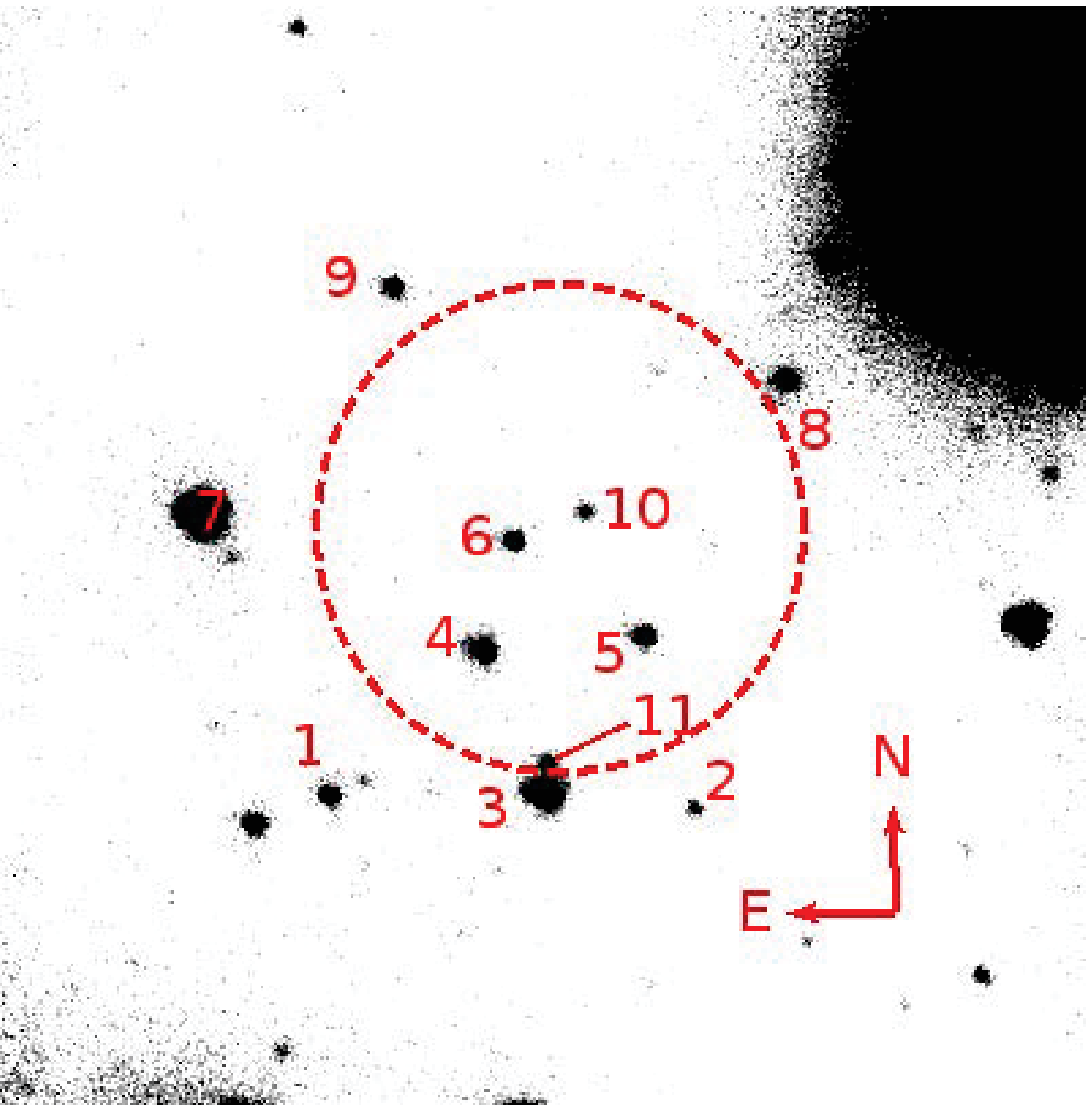}}
    \includegraphics[width=0.45\textwidth]{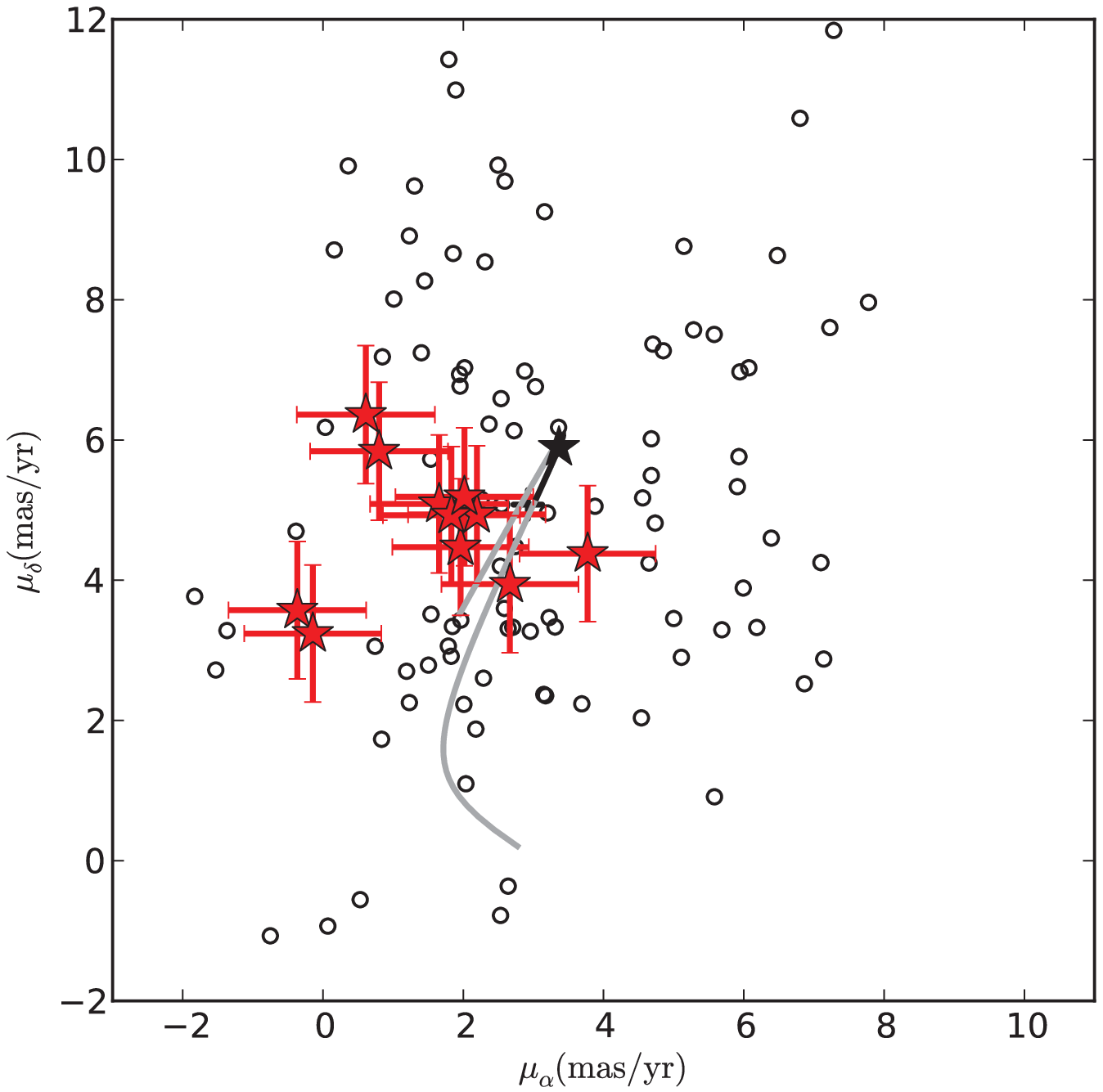}
    \caption{\textit{Left Panel:} A $4 \times 4^{\prime\prime}$ $K_s$ band image around the \textit{Chandra} position of AXP\,1E\,1841$-$045 (Wachter et al. 2004). The red circle is the 3-$\sigma$ position error ($0.9^{\prime\prime}$ radius). Sources are labelled as per Testa et al. (2008). None of the sources in the error circle are conclusively identified as the counterpart of the magnetar. \textit{Right Panel:} The proper motions of all the objects in the vicinity of AXP\,1E\,1841$-$045 are marked with red error bars. The other stars in the $10 \times 10^{\prime\prime}$ image are marked by black circles. The black star denotes the proper motion of a hypothetical progenitor at the distance of Kes\,73 moving with the Galactic rotation (thick gray line).} 
    \label{fig:1e1841}
  \end{center}
\end{figure}

The left panel of Figure~\ref{fig:1e1841} shows a $4 \times 4^{\prime\prime}$ cutout from our $K_s$ band image around the X-ray position of AXP\,1E\,1841$-$045 (red circle, Wachter et al. 2004). We registered our NIRC2 images to the 2MASS catalog with RMS residuals of 20\,milli-arcseconds. The accuracy of the 2MASS coordinates for the brightness of our registration stars ($\mathrm{K_s\,mag} \approx 11$) is $70 - 80$ milli-arcseconds. Testa et al. (2008) proposed star 9 as the counterpart for the AXP based on a 3-$\sigma$ photometric variability. However, our astrometry shows that it lies outside the \textit{Chandra} error circle. Given the NIR flux to X-ray flux ratios for other magnetars and the quiescent X-ray flux from AXP\,1E\,1841$-$045, it is highly probable that one of the objects in the field is the NIR counterpart of the magnetar.

Without an identified counterpart, we can only set upper limits to the proper motion of AXP\,1841$-$045. The right panel of Figure~\ref{fig:1e1841} shows the proper motions of all the objects in the vicinity of the AXP. The proper motions are corrected for the motion of the Milky Way as per Tendulkar et al. (2012). The upper limit for the proper motions of any of the objects is $\sim4$\,milli-arcseconds\,yr$^{-1}$. At the nominal distance of Kes\,73 (8.5\,kpc, Tian \& Leahy 2008), this corresponds to a transverse space velocity of $160\,\mathrm{km\,s^{-1}}$.

\section{Conclusions}
Our proper motion measurements are consistent with the radio VLBI proper motion measurements of AXP\,1E\,1810$-$197 by Helfand et al. (2007) and of PSR\,J1550$-$5418 by Deller et al. (2012). With these results in hand, the space velocities of magnetars are similar to the $\sim200-300\,\mathrm{km\,s^{-1}}$ velocities of pulsars (Hobbs et al. 2005). 



\end{document}